%% file: paper.tex
\documentclass[11pt, a4paper]{article}

\usepackage{fullpage}
\usepackage{latexsym}
\usepackage{psfig}

\input{macrosarticle.tex}

\begin{document}

\title{Pole Dynamics for Elliptic Solutions of the Korteweg-deVries Equation}

\author{Bernard Deconinck\\
Mathematical Sciences Research Institute\\
1000 Centennial Drive\\
Berkeley CA 94720\\
~\\
Harvey Segur\\
Department of Applied Mathematics\\
University of Colorado\\
Boulder, CO 80309-0526
}

\maketitle

\begin{abstract}

The real, nonsingular elliptic solutions of the Korteweg-deVries equation are
studied through the time dynamics of their poles in the complex plane.  The
dynamics of these poles is governed by a dynamical system with a constraint.
This constraint is shown to be solvable for any finite number of poles located
in the fundamental domain of the elliptic function, often in many different
ways. Special consideration is given to those elliptic solutions that have a
real nonsingular soliton limit.

\end{abstract}


\section{Introduction}

In 1974, Kruskal \cite{kruskalpoles} considered the interaction of solitons
governed by the Korteweg-deVries equation (KdV),

\beq\la{kdv}
u_t=6 u u_x+u_{xxx}.
\eeq

\no Each KdV soliton is defined by a meromorphic function in the complex 
$x$-plane ($i.e., \mbox{sech}^2k(x-x_0))$, so Kruskal
\cite{kruskalpoles} suggested that the interaction of two or more
solitons could be understood in terms of the dynamics of the poles of these
meromorphic functions in the complex x-plane, where the poles move according
to a force law deduced from \rf{kdv}. This was followed by the work of
Thickstun \cite{thickstun} who considered the case of two solitons in great
detail. 

Following a different line of thought, Airault, McKean and Moser \cite{amm}
studied rational and elliptic solutions of the KdV equation. An elliptic
solution of KdV is by definition a solution of the KdV equation that is
doubly periodic and meromorphic in the complex $x$-plane, for all time. Note
that the soliton case is an intermediary case between the elliptic and the
rational case. It was treated as such in \cite{amm}.

Airault, McKean and Moser \cite{amm} approached these elliptic solutions and
their degenerate limits through the motion of their poles $x_i(t)$ in the
complex $x$-plane. In particular, they looked for elliptic KdV solutions of the
form 

\beq\la{sol}
u(x,t)=-2\sum_{i=1}^N \wp(x-x_i(t)).
\eeq

\no Here $\wp(z)$ denotes the \weier elliptic function. It can be defined
by its meromorphic expansion 

\beq\la{wp}
\wp(z)=\frac{1}{z^2}+\sum_{(m,n)\neq (0,0)}\left(\frac{1}{(z+2 m \omega_1+2 n
\omega_2)^2}-\frac{1}{(2 m \omega_1+2 n \omega_2)^2}\right),
\eeq

\no with $\omega_1/\omega_2$ not real\footnote{In this paper it is always
assumed that $\omega_1$ is real and $\omega_2$ is imaginary. This is necessary
to ensure reality of the KdV solution \rf{sol} when $x$ is restricted to the
real $x$-axis. Other considerations for reality of the elliptic KdV solutions
will be discussed in Section \ref{sec:dis}.}. More properties of the \weier
function will be given as they are needed.  

It is shown in \cite{amm} that the dynamics of the poles $x_i(t)$ is governed
by the dynamical system

\alpheqn

\beq\la{dynsys}
\dot{x_i}=12 \sum_{j=1, j\neq i}^N\wp(x_i-x_j),~~~i=1,2,\ldots, N,
\eeq

\no (the dot denotes differentiation with respect to time) with the invariant
constraint

\beq\la{constraint}
\sum_{j=1, j\neq i}^N \wp'(x_i-x_j)=0.~~~i=1,2,\ldots, N.
\eeq

\resetalpheqn

\no Here the prime denotes differentiation with respect to the argument.  The
solutions \rf{sol} generalize an elliptic solution given earlier by Dubrovin
and Novikov \cite{dubnov}, corresponding to the case $N=3$. These authors also
recall the Lam\'{e}-Ince potentials \cite{ince2}

\beq\la{lame}
u(x)=-g(g+1)\wp(x),
\eeq

\no which are the simplest $g$-gap potentials of the stationary Schr\"{o}dinger
equation

\beq\la{schrodinger}
\ppn{2}{\psi}{x}+u(x) \psi=\lambda \psi.
\eeq

\no The remarkable connection between the KdV equation and the stationary
Schr\"{o}dinger equation has been known since the work of Gardner, Greene,
Kruskal and Miura \cite{ggkm}. Dubrovin and Novikov show \cite{dubnov} that
the $(N=3)$-solution discussed in \cite{dubnov} is a 2-gap solution of the KdV
equation with a 2-gap Lam\'e-Ince potential as initial condition. 

If one considers the rational limit of the solution \rf{sol} ($i.e.$, the
limit in which the \weier function $\wp(z)$ reduces to $1/z^2$), then the
constraint \rf{constraint} is only solvable for a triangular number of poles, 

\beq\la{triangular}
N=\frac{n(n+1)}{2},
\eeq

\no for any positive integer $n$ \cite{amm}. Notice that the Lam\'{e}-Ince
potentials are given by $g(g+1)/2$ times an $N=1$ potential. Based on these
observations, it was conjectured in \cite{amm} that also in the elliptic case
given by \rf{sol}, the constraint \rf{constraint} is only solvable for a
triangular number $N$, ``or very nearly so''. From the moment it appeared this
conjecture was known not to hold, because it already fails in the soliton
case, where the \weier function degenerates to hyperbolic functions. This
failure of the conjecture easily follows from the work of Thickstun
\cite{thickstun}.  

A further understanding of the elliptic case had to wait until 1988, when 
Verdier provided more explicit examples of elliptic potentials of the
Schr\"{o}dinger equation \cite{verdier}. Subsequently, Treibich and Verdier
demonstrated that 

\beq\la{tv}
u(x)=-2 \sum_{i=1}^4 \frac{g_i(g_i+1)}{2} \wp (x-x_0-\omega_i)
\eeq

\no ($\omega_3=\omega_1+\omega_2, \omega_4=0$, the $g_i$ are positive
integers) are finite-gap potentials of the stationary Schr\"{o}dinger equation
\rf{schrodinger}, and hence result in elliptic solutions of the KdV equation
\cite{tv1,tv2,tv3}. 

The potentials of Treibich and Verdier were generalized by Gesztesy and
Weikard \cite{gw1,gw2}. They showed that any elliptic finite-gap potential of
the stationary Schr\"{o}dinger equation \rf{schrodinger} can be represented in
the form 

\beq\la{gw}
u(x)=-2 \sum_{i=1}^M \frac{g_i(g_i+1)}{2} \wp(x-\alpha_i),
\eeq

\no for some $M$ and positive integers $g_i$. Notice that this formula
coincides with \rf{sol} if all the $g_i$ are 1. 

The focus of this paper is the constrained dynamical system (\ref{dynsys}-b).
We return to the ideas put forth by Kruskal \cite{kruskalpoles} and Thickstun
\cite{thickstun}. This allows us to derive the system (\ref{dynsys}-b) in a
context which is more general than \cite{amm}: there it was obtained as a
system describing a class of special solutions of the KdV equation. Here, it
is shown that any meromorphic solution of the KdV equation which is doubly
periodic in $x$ is of the form \rf{sol}. Hence the consideration of solutions
of the form \rf{sol} and the system of equations
(\ref{dynsys}-\ref{constraint}) leads to {\em all} elliptic solutions of the
KdV equation.   

Simultaneously, some of the results of Gesztesy and Weikard \cite{gw1, gw2}
are recovered here. Because of the connection between the KdV equation and the
Schr\"{o}dinger equation, any potential of Gesztesy and Weikard can be used as
an initial condition for the KdV equation, which determines any time
dependence of the parameters of the elliptic KdV solution with that initial
condition. Our approach demonstrates which parameters in the solutions \rf{gw}
are time independent and which are time dependent. 

The following conclusions are obtained in this paper:

\begin{itemize}

\item All finite-gap\footnote{See Section \ref{sec:dynsys}} elliptic solutions
of the KdV equation are of the form \rf{sol}, for almost all times (see
below).  In other words, if u(x,t) is a finite-gap KdV solution that is doubly
periodic in the complex x-plane, then u necessarily has the form \rf{sol}
except at isolated instants of time.

\item Any number $N \neq 2$ of $x_j$ is allowed in \rf{sol}, if the ratio
$|\omega_1/\omega_2|\gg 1$. It follows that in this case, the constraint
\rf{constraint} is solvable for any positive integer\footnote{The constraint
\rf{constraint} is solvable for $N=2$ \cite{amm}. As noted there, the
corresponding solution reduces to a solution for $N=1$, with smaller periods.
This case is therefore trivial and is disregarded} $N\neq 2$. Our method for
demonstrating this is to use a deformation from the soliton limit of the
system (\ref{dynsys}-b). Furthermore, this method is useful because it
provides good initial guesses for the numerical solution of \rf{constraint} in
Section \ref{sec:ex}. This method does not recover all elliptic solutions of
the KdV equations, because only those solutions are found which have
nonsingular soliton limits. In particular, the solutions corresponding to the
Treibich-Verdier potentials \rf{tv} are not found. 

\item If $|\omega_1/\omega_2|\gg 1$, then for a given $N>4$, nonequivalent
configurations satisfying the constraint exist that do {\em not} flow into
each other under the KdV flow and which cannot be translated into each other.
To the best of our knowledge this is a new result. 

\item The $x_i$ are allowed to coincide, but only in triangular numbers: if
some of the $x_i$ coincide at a certain time $t_c$, then $g_i(g_i+1)/2$ of
them coincide at that time $t_c$. At this time $t_c$, the solution can be
represented in the form \rf{gw} with not all $g_i=1$.

Such times $t_c$ are referred to as collision times and the poles are said to
collide at the collision time. Before and after each collision time all $x_i$
are distinct, hence pole collisions are isolated events. At the collision
times, the dynamical system \rf{dynsys} is not valid. The dynamics of the
poles at the collision times is easily determined directly from the KdV
equation. 

Gesztesy and Weikard \cite{gw1, gw2} demonstrate that \rf{gw} are elliptic
potentials of the Schr\"{o}dinger equation. These potentials generalize to
solutions of the KdV equation, but this requires the $g_i$ to be nonsmooth
functions of time. Only at the collision times $t_c$ are the $g_i$ not all
one. Furthermore, at all times but the collision times, the number of
parameters $\alpha_i$ (which are time dependent) is $N=\sum_{i=1}^M
g_i(g_i+1)/2$. This shows that the generalization from \cite{gw1, gw2} to
solutions of the KdV equation is nontrivial. 

Using the terminology of Chapter 7 of \cite{belokolos1}, the poles are in {\em
general position} if all $g_i$ are equal to one. Otherwise, if not all
$g_i=1$, the poles are said to be in {\em special position}. We conclude that
at the collision times the poles are in special position. Otherwise they are
in general position. 

\item The solutions discussed here are finite-gap potentials of the stationary
Schr\"{o}dinger equation \rf{schrodinger}, with $t$ treated as a parameter. 
In other words, each solution specifies a one-parameter family of finite-gap
potentials of the stationary Schr\"{o}dinger equation. It follows from our
methods that to obtain a $g$-gap potential that corresponds to a nonsingular
soliton potential, one needs at least $N=g(g+1)/2$ poles $x_j$. The
Lam\'{e}-Ince potentials show that this lower bound is sharp. If we consider
potentials that do not have nonsingular soliton limits (such as the
Treibich-Verdier potentials \rf{tv}) then it may be possible to violate this
lower bound. 

\item The relationship between elliptic solutions and soliton solutions of
(\ref{dynsys}-b) and hence of the KdV equation is made explicit. In fact, this
relationship is essential to the method used here, as mentioned above. 

\item In Section \rf{sec:ex}, we present an explicit solution of the form
\rf{sol} with $N=4$. Notice that if in \rf{tv} all $g_i$ are one, the solution
is reducible to a solution with $N=1$ and smaller periods. To see this it is
convenient to draw the pole configuration corresponding to \rf{tv} in the
complex plane. This is actually true, even if $g_i$ is not one, but all $g_i$
are equal. In that case, \rf{tv} reduces to a Lam\'{e}-Ince potential
\rf{lame}. Unlike any of the Treibich-Verdier solutions, the $(N=4)$-solution,
presented in Section \rf{sec:ex}, has a nonsingular soliton limit. 

\end{itemize}

The first five conclusions are all discussed in Section \ref{sec:dis}. In
Section \ref{sec:thick}, the results of Kruskal \cite{kruskalpoles} and
Thickstun \cite{thickstun} for the soliton solutions of the KdV equation are
reviewed, but they are obtained from a point of view which is closer to the
approach we present in Section \ref{sec:dynsys} for the elliptic solutions of
the KdV equation. Finally, in Section \ref{sec:ex}, some explicit examples are
given, including illustrations of the motion of the poles in the complex
$x$-plane. 

\section{The soliton case: hyperbolic functions}\la{sec:thick}

In this section, the results of Kruskal \cite{kruskalpoles} and Thickstun
\cite{thickstun} for the dynamics of poles of soliton solutions are discussed
from a point of view that will allow us to generalize to the periodic case.  

Consider the one-soliton solution of the KdV equation

\beq\la{1sol}
u(x,t)=2 k^2 \mbox{sech}^2 k(x+4 k^2 t-\varphi). 
\eeq

\no Here $k$ is a positive parameter (the wave number of the soliton)
determining the speed and the amplitude of the one-soliton solution. Using the
meromorphic expansion \cite{tables}

\beq\la{merosol} \frac{1}{T^2}\mbox{cosech}^2
\frac{x}{T}=\sum_{n=-\infty}^{\infty}\frac{1}{(x+ i n \pi T)^2}, \eeq

\no (uniformly convergent except at the points $x=i n \pi T$) one easily
obtains the following meromorphic expansion for the one-soliton solution of
the KdV equation:

\beq\la{mero1sol}
u(x,t)=-2 k^2 \sum_{n=-\infty}^{\infty} \frac{1}{(k (x+4 k^2 t-\varphi)+i
\frac{\pi}{2}+i n \pi)^2}.
\eeq

\no From this expression, one easily finds that the locations of the poles of
the one-soliton solution of the KdV equation for all time are given by 

\beq\la{poles1sol}
x_n=\varphi-4 k^2 t-\frac{i \pi}{k}\left(n+\frac{1}{2}\right).
\eeq

\no This motion is illustrated in Fig. \ref{fig1}a. Notice that the locations
of the poles are symmetric with respect to the real $x$-axis. This is a
consequence of the reality of the solution \rf{1sol}. In order for a solution
to be real it is necessary and sufficient that if $x_n(t)$ is a pole, then so
is $x_n^*(t)$, where $~^*$ denotes the complex conjugate\footnote{The vertical
line of poles can be rotated arbitrarily. The expression \rf{mero1sol} still
results in a solution of the KdV equation, but it is no longer real. Again, we
will only consider real, nonsingular solutions, when restricted to the real
$x$-axis}. The closest distance between any two poles is $d=\pi/k$ and is
constant both along the vertical line $\mbox{Re}(x)=\varphi-4 k^2 t$ and in
time. Note that the poles are moving to the left. This is a consequence of the
form of the KdV equation \rf{kdv}, which has time reversed, compared to the
version Kruskal \cite{kruskalpoles} and Thickstun \cite{thickstun} used. 

\begin{figure}[htb]
\begin{tabular}{ccc}
\psfig{file=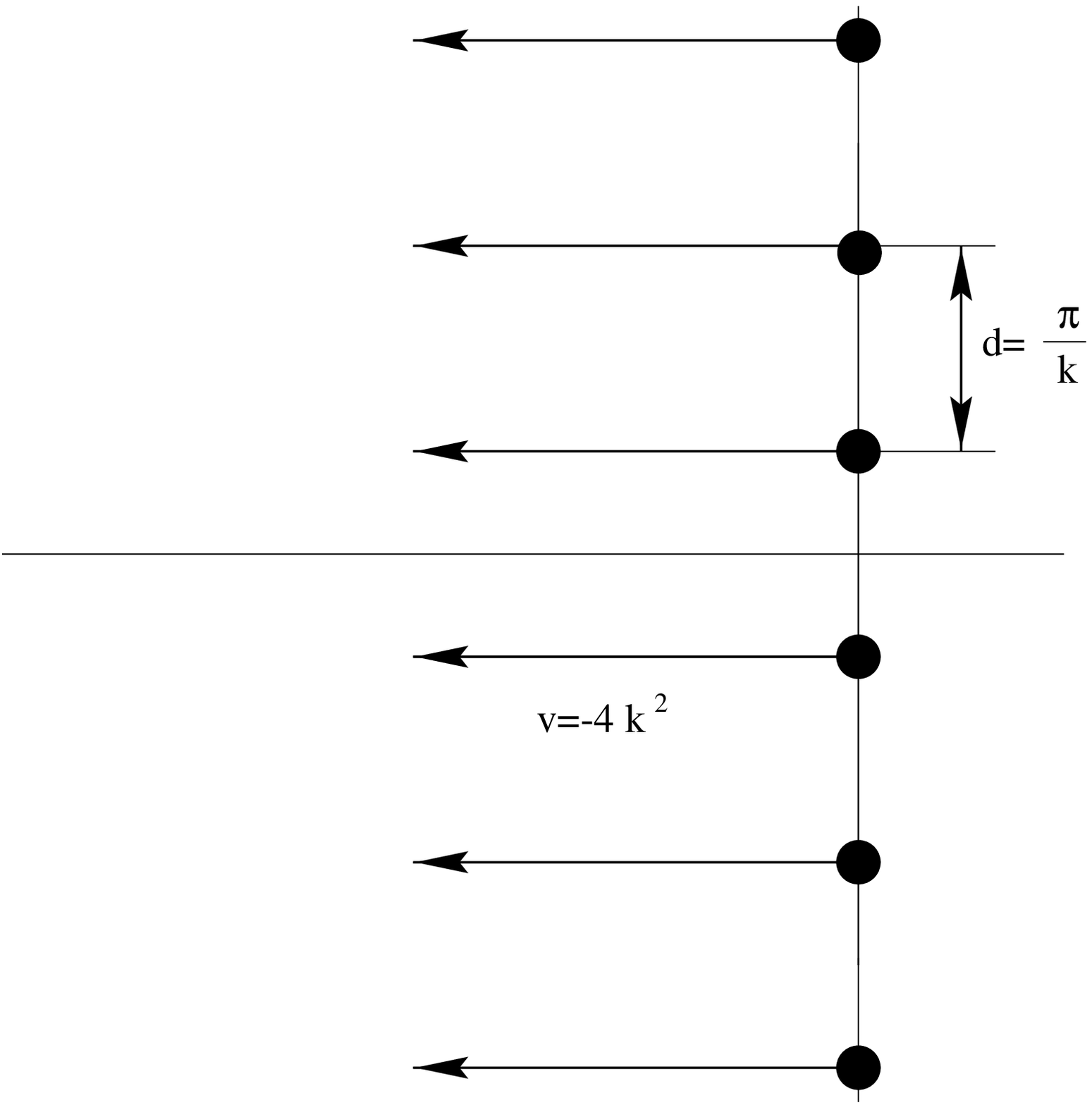,width=2.9in}
&~~~&
\psfig{file=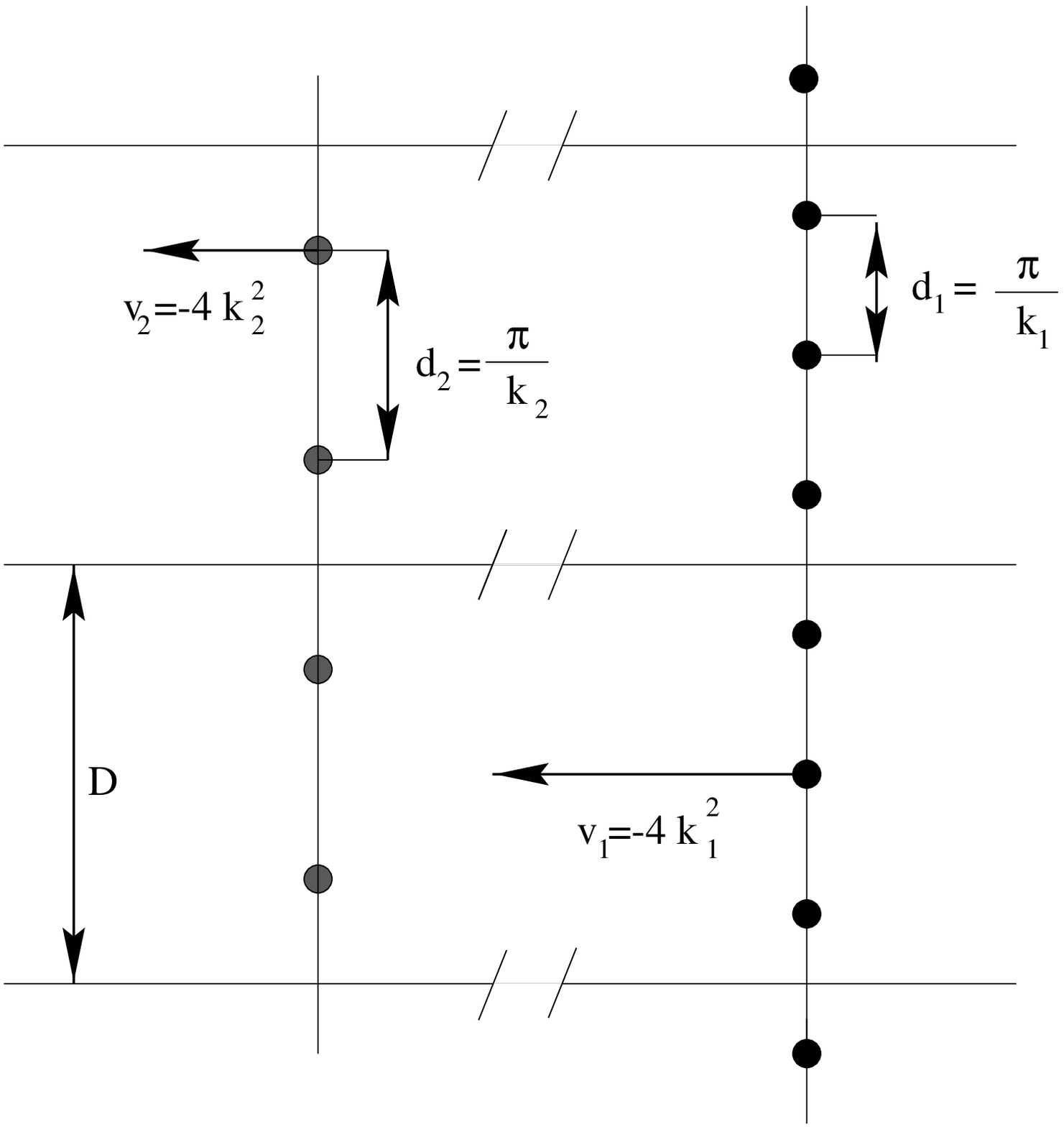,width=2.9in}\\
(a)&&(b)
\end{tabular}
\caption{\la{fig1}{\bf (a) The motion of the poles of a one-soliton solution in
the complex plane. (b) The asymptotic motion of the poles of a two-soliton
solution.}}
\end{figure}

Since two solitons of the KdV equation cannot move with the same speed, a
two-soliton solution of the KdV equation asymtotically appears as the sum of
two one-soliton solutions which are well-separated: the higher-amplitude
soliton, which is faster, is to the right of the smaller-amplitude soliton as
$t\rightarrow -\infty$. As $t\rightarrow \infty$, the higher-amplitude soliton
is to the left of the smaller-amplitude soliton. Hence as $t\rightarrow
-\infty$, the pole configuration of a two-soliton solution with wave numbers
$k_1$ and $k_2$ is as in Fig. \ref{fig1}b. In this limit, the two-soliton
solution is a sum of two one-soliton solutions. Each results in a vertical
line of equispaced poles, with interpolar distance respectively $d_1=\pi/k_1$
and $d_2=\pi/k_2$. As long as the solitons are well-separated, these poles
move in approximately straight lines, parallel to the real axis, with
respective velocities $v_1=-4 k_1^2$ and $v_2=-4 k_2^2$. Since $|v_1|>|v_2|$,
the solitons interact eventually. This interacting results in non-straight
line motion of the poles. After the interaction, the situation is as in Fig.
\ref{fig1}b, but with the two lines of poles interchanged. 

Thickstun \cite{thickstun} considered the case where $k_1$ and $k_2$ are
rationally related, so $k_1/k_2=p/q$, where $p$ and $q$ are positive integers.
In this case, one can define $D=p d_1=q d_2$. The complex $x$-plane is now
divided into an infinite number of equal strips, parallel to the real $x$-axis,
each of height $D$. The real $x$-axis is usually taken to be the base of such
a strip. It is easy to show \cite{thickstun} that the motion of the poles in
one strip is repeated in every strip. Hence, one is left studying the motion
of a finite number $N$ ($=p+q$) of poles in the fundamental strip, whose base
is the real $x$-axis. Thickstun examined this motion by analyzing the exact
expression for a two-soliton solution of the KdV equation. 

Any two-soliton solution is expressible as \cite{as}

\beq\la{twosol}
u(x,t)=2 \p_x^2 \ln \tau(x,t).
\eeq

\no It follows from this formula that the poles of $u(x,t)$ are the zeros of
$\tau(x,t)$ if $\tau(x,t)$ is entire in $x$. Then the \weier Factorization
Theorem \cite{conway} gives a factorization for $\tau(x,t)$:

\beq\la{factau}
\tau(x,t)=C \prod_{k=1}^\infty \left(1-\frac{x}{x_k}\right)e^{x/x_k}.
\eeq

\no Since only the second logarithmic derivative of this function is relevant,
the constant $C$ is not important. If the solution is periodic in the
imaginary $x$-direction, this is rewritten as

\beq\la{factau2} \tau(x,t)=C
\prod_{n=1}^{N}\prod_{l=-\infty}^{\infty}\left(1-\frac{x}{x_n+i l
D}\right)e^{x/(x_n+i l D)}, \eeq

\no where the first product runs over the poles in the fundamental strip. The
second product runs over all strips. Using the uniform convergence of
\rf{factau2}, 

\beq\la{usol}
u(x,t)=-2 \sum_{n=1}^{N}\sum_{l=-\infty}^{\infty}\frac{1}{(x-x_n-i l D)^2},
\eeq

\no which, using \rf{merosol}, is rewritten as 

\beq\la{usolhyp} u(x,t)=-2
\frac{\pi^2}{D^2}\sum_{n=1}^{N}\mbox{cosech}^2\frac{\pi(x-x_n)}{D}
=-2 \frac{k_2^2}{q^2}\sum_{n=1}^{p+q}\mbox{cosech}^2 \frac{k_2 (x-x_n)}{q}
, \eeq

\no where the pole locations $x_n$ depend on time: $x_n=x_n(t)$. One recovers
the one-soliton solution \rf{1sol} easily, by equating $k_1=0, p=0,q=1$.
Equation \rf{usolhyp} essentially expresses a two-soliton solution as a linear
superposition of $N$ one-soliton solutions with nonlinearly interacting
phases. Note that the first equality in \rf{usolhyp} is valid for arbitrary
soliton solutions that are periodic in $x$ with period $i D$. This is the case
for a g-soliton solution if its wavenumbers $k_i$, $i=1,2,\ldots, g$ are all
commensurable: $(k_1:k_2:\ldots:k_g)=(p_1:p_2:\ldots:p_g)$, for positive
distinct integers $p_i$, $i=1,2,\ldots, g$ which have no overall common
integer factor. The total number of poles in a strip is then
$N=p_1+p_2+\ldots+p_g$. In obtaining \rf{usol} and \rf{usolhyp}, we have
deviated from Thickstun's approach \cite{thickstun} to an approach that is
generalized to the elliptic case of the next section in a straightforward way.

Next, we derive the dynamics imposed on the poles $x_n(t)$ by the KdV
equation. This is conveniently done by substituting \rf{usol} into \rf{kdv} and
examining the behaviour near one of the poles: $x=x_n+\epsilon$. This results
in several singular terms as $\epsilon\rightarrow 0$, corresponding to
negative powers of $\epsilon$. The dynamics of the poles is then determined by
the vanishing of the coefficients of these negative powers and the zeroth
power. This results in only two nontrivial equations, obtained at order
$\epsilon^{-3}$ and $\epsilon^{-2}$ respectively:

\alpheqn

\bea \dot{x}_n&=&12 {\sum_{k=1, k\neq n}^N}\sum_{l=-\infty}^\infty
\frac{1}{(x_k-x_n-i l D)^2}+12 \sum_{l\neq 0, l=-\infty}^{\infty}\frac{1}{(-i
l D)^2},\\ 0&=&\sum_{k=1, k\neq n}^N \sum_{l=-\infty}^\infty
\frac{1}{(x_k-x_n-i l D)^3}, \eea

\resetalpheqn

\no for $n=1,2,\ldots, N$. Using \rf{merosol} and its derivative,

\alpheqn

\bea\la{soldynsys}
\dot{x}_n&=&-4 \frac{\pi^2}{D^2}+12\frac{\pi^2}{D^2}
{\sum_{k=1, k\neq n}^N} \mbox{cosech}^2 \frac{\pi (x_k-x_n)}{D},\\\la{solcons}
0&=&\sum_{k=1, k\neq n}^N \mbox{cosech}^2 \frac{\pi (x_k-x_n)}{D}
\,\mbox{coth}\frac{\pi (x_k-x_n)}{D},
\eea

\resetalpheqn

\no for $n=1,2,\ldots, N$. Hence the dynamics of the poles $x_n(t)$ is
determined by \rf{soldynsys}. This dynamics is constrained by the equations
\rf{solcons}. These constraint equations \rf{solcons} are invariant under the
flow of \rf{soldynsys}. This follows from a direct calculation. 

\vspace*{12pt}

\no {\bf Remarks:}

\begin{itemize}

\item Since the KdV equation has two-soliton solutions for any ratio of the
wavenumbers $k_1/k_2\neq 1$, the constraint \rf{solcons} is solvable for any
value of $N$, excluding $N=2$, which can only be obtained by $p=q=1$,
resulting in equal wavenumbers $k_1$ and $k_2$. 

\item In particular, it follows that the minimum number of poles in a
fundamental strip required to obtain a $g-soliton$ solution is
$N=1+2+3+\ldots+g=g(g+1)/2$, corresponding to a $g$-soliton solution with
wavenumbers which are related as
$(k_1:k_2:\ldots:k_{g-1}:k_g)=(g:g-1:\ldots:2:1)$. 

\item Equating $k_1=0$, $p=0$, $q=1$, one obtains from \rf{soldynsys}
$\dot{x}_1=-4 k_2^2$, corresponding to the dynamics of the one-soliton
solution. The asymptotic behavior of the poles of a two-soliton solution also
follows from \rf{soldynsys}: from the separation of the poles into distinct
vertical lines, it follows from \rf{dynsys} that the velocity of these
vertical lines is given by the one-soliton velocity for each line, as
expected. This result follows from easy algebraic manipulation and the
identity
\beq\nonumber
\frac{p^2-1}{3}=\sum_{n=1}^{p-1}\mbox{cosec}^2\left(\frac{n \pi}{p}\right),
\eeq
\no valid for any integer $p>1$. 

\item A full analysis of the interaction of the poles for the case of any
two-soliton solution with $k_1/k_2=p/q$ is given in \cite{thickstun}. 

\end{itemize}

\section{The elliptic case}\la{sec:dynsys}

Consider the quasiperiodic finite-gap solutions of the KdV equation with $g$
phases \cite{itsmatveev}

\beq\la{thetasol}
u(x,t)=2 \p_x^2 \ln \theta_g(\mbf{k}x+\mbf{\omega} t+\mbf{\phi}|\mbf{B}),
\eeq

\no where 

\beq \theta_g(\mbf{z}|\mbf{B})=\sum_{\mbf{m}\in\mbf{Z^g}}
\exp\left(\frac{1}{2}\mbf{m}\cdot\mbf{B}\cdot\mbf{m}+i
\mbf{m}\cdot\mbf{z}\right), \eeq

\no a hyperelliptic Riemann-theta function of genus $g$. The $g\times g$
real Riemann matrix ($i.e.$, symmetric and negative definite)
$\mbf{B}$ originates from a hyperelliptic Riemann surface with only one point
at infinity. Furthermore, $\mbf{k}$, $\mbf{\omega}$ and $\mbf{\phi}$ are
$g$-dimensional vectors.

The derivation of equations \rf{usolhyp}, \rf{soldynsys} and \rf{solcons} is
easily generalized to the case where the solution is not only periodic in the
imaginary $x$-direction, but also in the real $x$-direction:

\beq\la{uper}
u(x+L_1,t)=u(x,t)=u(x+i L_2,t).
\eeq

\no This divides the complex $x$-plane into an array of rectangular domains,
each of size $L_1\times L_2$. One of these domains, called the fundamental
domain $S$, is conveniently placed in the lower left corner of the first
quadrant of the $x$-plane. The theta function has the property \cite{dub}

\beq\la{thetaprop}
\theta_g(\mbf{z}+i \mbf{B}\cdot \mbf{M}+2 \pi \mbf{N}|\mbf{B})=
\theta_g(\mbf{z}|\mbf{B})\exp\left(-\frac{1}{2}\mbf{M}\cdot \mbf{B}
\cdot \mbf{M}
+i \mbf{M}\cdot \mbf{z}\right),
\eeq

\no for any pair of $g$-component integer vectors $\mbf{M}, \mbf{N}$. This
expression is useful to determine conditions on the wavevector $\mbf{k}$ in
order for $u(x,t)$, given by \rf{thetasol}, to satisfy \rf{uper}: 

\beq\la{conper}
\exists~ \mbf{N_0}, \mbf{M_0} \in \mbf{Z}^g:\mbf{k} L_1=2 \pi \mbf{N_0}, 
~ \mbf{k} L_2=\mbf{B}\cdot \mbf{M_0}.
\eeq

\no These results are now used to determine the number of poles $N$ of
$u(x,t)$ in the fundamental domain. The poles of \rf{thetasol} are given by
the zeros of $\vartheta(x,t)=\theta_g(\mbf{k}x+\mbf{\omega}
t+\mbf{\phi}|\mbf{B})$, regarded as a function of $x$:

\bea\nonumber
N&=&\frac{1}{2 \pi i} \oint_{\p S} d \ln \vartheta(x,t)\\\nonumber
&=&\frac{1}{2 \pi i}\left(
\int_{0}^{L_1}d \ln \vartheta(x,t)+\int_{0}^{i L_2} d \ln
\vartheta(x+L_1,t)-\right.\\\nonumber
&&\left.\int_{0}^{L_1} d \ln \vartheta(x+i L_2,t)-\int_{0}^{i L_2} d \ln
\vartheta(x,t)
\right)\\\nonumber
&\since{thetaprop}&\frac{1}{2 \pi i}\int_{0}^{L_1} d \ln
\frac{\vartheta(x,t)}{\vartheta(x+i L_2,t)}\\\nonumber
&\since{thetaprop}&\frac{1}{2 \pi i} \int_{0}^{L_1} d (-i x \mbf{M_0}\cdot
\mbf{k})\\\la{numberofpoles}
&=&-\mbf{M_0}\cdot\mbf{N_0}
=-\frac{L_1}{2 \pi L_2}\mbf{M_0}\cdot\mbf{B}\cdot\mbf{M_0}.
\eea

\no The first equality of \rf{numberofpoles} confirms that $N$ is an integer.
The second equality shows that $N$ is positive, by the negative-definiteness
of $\mbf{B}$. 

We now proceed to determine the dynamical system satisfied by the motion of
the $N$ poles of $u(x,t)$ in the fundamental domain $S$. Again, the poles of
$u(x,t)$ are the zeros of $\vartheta(x,t)$. Furthermore, simple zeros of
$\vartheta(x,t)$ result in double poles of $u(x,t)$, as in the hyperbolic
case. 

The \weier Factorization theorem \cite{conway} gives the following form for
$\vartheta(x,t)$:

\beq\la{factheta} \vartheta(x,t)=e^{c
x^2/2}\prod_{k}\left(1-\frac{x}{x_k}\right)e^{\frac{x}{x_k}+\frac{x^2}{2
x_k^2}}, \eeq

\no where the product runs over all poles $x_k$. The additional exponential
factors, as compared to \rf{factau}, are required because the poles now appear
in a bi-infinite sequence: both in the vertical and horizontal directions.
These exponential factors ensure uniform convergence of the product. The
parameter $c$ is allowed to depend on time. It determines the behavior of
$\vartheta(x,t)$ as $x$ approaches infinity in the complex $x$-plane \cite{dub}. Using
\rf{uper}, this is rewritten as 

\bea\nonumber
\vartheta(x,t)=\exp(cx^2/2) \prod_{n=1}^N \prod_{m=-\infty}^\infty
\prod_{l=-\infty}^\infty &&\!\!\!\!\!\!\!\!\!\!\left(1-\frac{x}{x_n+m L_1+i l
L_2}\right)\times\\\la{facthetaper}
&&\!\!\!\!\!\!\!\!\!\!
\exp\left(\frac{x}{x_n+m L_1+i l L_2}+
\frac{x^2}{2(x_n+m L_1+i l L_2)^2}\right).
\eea

\no The first product runs over the number of poles ($N$) in the fundamental
domain, the second and third products result in all translations of the
fundamental domain. From the uniform convergence of \rf{facthetaper}, 

\beq\la{u1} u(x,t)=2 c-2 \sum_{n=1}^N
\sum_{m=-\infty}^{\infty}\sum_{l=-\infty}^{\infty} \left(\frac{1}{(x-x_n-m
L_1-i l L_2)^2}-\frac{1}{(x_n+m L_1+i l L_2)^2}\right).  \eeq

\no Using the definition of the \weier function \rf{wp}, this is rewritten
as 

\beq\la{u2}
u(x,t)=2 c-2 \sum_{n=1}^N \wp(x-x_n)+2 \sum_{j=1}^N\wp(x_n), 
\eeq

\no where the periods of the \weier function are given by $2 \omega_1=L_1, 2
\omega_2=i L_2$. Define

\beq\la{constant}
\tilde{c}=2 c+2 \sum_{n=1}^N\wp(x_n).
\eeq

\no The dynamics of the poles $x_n=x_n(t)$ is determined by substitution of
\rf{u2} or \rf{u1} into the KdV equation and expanding in powers of $\epsilon$
for $x$ near a pole: $x=x_k+\epsilon$. Equating the coefficients of
$\epsilon^{-3}$, $\epsilon^{-2}$ and $\epsilon^0$ to zero result in 

\alpheqn

\bea\la{eldynsys}
\dot{x}_n&=&12 \sum_{j=1, j\neq n}^N \wp(x_j-x_n),\\\la{elcons}
0&=&\sum_{j=1, j\neq n}^N \wp'(x_j-x_n),\\
\dot{\tilde{c}}&=&0 \iff \tilde{c}(t)=\alpha=0,
\eea

\resetalpheqn

\no for $n=1,2,\ldots, N$. (The constant $\alpha$ can always be removed by a
Galilean shift, so it is equated to zero, without loss of generality.) The
constraints \rf{elcons} are invariant under the flow, as can be checked by
direct calculation. Notice that (\ref{eldynsys}-b) are identical to the
equations obtained by Airault, McKean and Moser \cite{amm}. These equations
are obtained here in greater generality: any solution \rf{thetasol} which is
doubly periodic in the $x$-plane gives rise to a system
(\ref{eldynsys}-b). This allows us to reach the conclusions stated
in the next section. 

\vspace*{12pt}

\no {\bf Remarks}

\begin{itemize}

\item In the limit $L_1\rightarrow \infty$, the equations
(\ref{eldynsys}--\ref{elcons}) reduce to (\ref{soldynsys}--\ref{solcons}). This
limit is most conveniently obtained from the Poisson representation of the
\weier function:
\beq\la{weierpois}
\wp(x)=\left(\frac{\pi}{L_2}\right)^2
\left(\frac{1}{3}+\mbox{cosech}^2\frac{\pi x}{L_2}+\sum_{n=-\infty, n\neq
0}^{\infty}\left\{\mbox{cosech}^2 \frac{\pi}{L_2}(x+n
L_1)-\mbox{cosech}^2\frac{n \pi L_1}{L_2}\right\}\right).
\eeq

\no This representation is obtained from \rf{wp} by working out the summation
in the vertical direction. It gives the \weier function as a sum of
exponentially localized terms, hence few terms have important contributions in
the fundamental domain. A Poisson expansion for $\wp'(x)$ is obtained from
differentiating \rf{weierpois} term by term with respect to $x$. 

\item Define the one-phase theta function $\theta_1(z,q)$ \cite{tables}:

\beq\la{theta1}
\theta_1(z,q)=2 \sum_{n=0}^{\infty}(-1)^n q^{(n+1/2)^2}\sin(2n+1)z,
\eeq

\no with $|q|<1$. If $L_2<L_1$, then the relationship $\wp(z)=a-\p_x^2 \ln
\theta_1(\pi z/L_1, i L_2/L_1)$, with $a$ a constant \cite{tables}, allows us
to rewrite \rf{sol} as

\beq\la{connysol}
u(x,t)=\hat{a}+2 \p_x^2 \ln \prod_{j=1}^N \theta_1\left(\frac{\pi}{L_1}
(x-x_j(t)),i \frac{L_2}{L_1}\right),
\eeq

\no with $\hat{a}=-2 a N$. Hence, for the doubly-periodic solutions of the KdV
equation of the form \rf{thetasol}, it is possible to rewrite the $g$-phase
theta function as a product of $N$ $1$-phase theta functions, with nonlinearly
interacting phases. Note that this does not imply that the $g$-phase theta
function appearing in \rf{thetasol} is reducible. Reducible theta-functions do
not give rise to solutions of the KdV equation \cite{dub}.

\item By taking another time derivative of \rf{eldynsys} and using \rf{elcons},
one obtains 
\beq\la{elsecond}
\ddot{x}_n=-(12)^2 \sum_{j=1, j\neq n}^N \wp'(x_j-x_n)\wp(x_j-x_n).
\eeq

\no It is known that this system of differential equations is Hamiltonian
\cite{chud}, with Hamiltonian

\beq\la{hamil1}
H=\frac{1}{2}\sum_{k=1}^{N}p_k^2+\frac{(12)^2}{2}\sum_{k=1}^N 
\sum_{j=1, j\neq k}^N \wp^2(x_k-x_j),
\eeq

\no and canonical variables $\{x_k,p_k=\dot{x}_k\}$. A second Hamiltonian
structure for the equations \rf{elsecond} is given in \cite{chud}. A Lax
representation for the system (\ref{eldynsys}-c) is also given there. This Lax
representation is a direct consequence of the law of addition of the \weier
function \cite{tables}. It is unknown to us whether a Hamiltonian structure 
exists for the constrained first-order dynamical system (\ref{eldynsys}-b). 

The Hamiltonian structure \rf{hamil1} shows that the system (\ref{eldynsys}-b)
is a (constrained) member of the elliptic Calogero-Moser hierarchy
\cite{belokolos1}.

\end{itemize}

\section{Discussion of the dynamics}\la{sec:dis}

In this section, the constrained dynamical system (\ref{eldynsys}-c) is
discussed. In particular, the assertions made in the introduction are
validated here. 

For reality of the KdV solution \rf{thetasol} when $x$ is restricted to the
real line, it is necessary and sufficient that if $x_j(t)$ appears, then so
does $x_j^*(t)$. Because the \weier function is a meromorphic functions of its
argument, this reality constraint is invariant under the dynamics
\rf{eldynsys}. 

As a consequence, the distribution of the poles in the fundamental domain $S$
is symmetric with respect to the horizontal centerline of $S$. Poles are
allowed on the centerline. Most of what follows is valid for both real KdV
solutions\footnote{``Real KdV solution'' refers to a solution of the KdV
equation which is real when $x$ is restricted to the real $x$-axis} and KdV
solutions that are not real, but we restrict our attention to real KdV
solutions. 

\subsection{All finite-gap elliptic solutions of the KdV equation are of the
form \rf{sol}, up to a constant}

A straightforward singularity analysis of the KdV equation \cite{amm} shows
that any algebraic singularity of a solution of the KdV equation is of the
type $u(x,t)=-2/(x-\alpha(t))^2+{\cal O}(x-\alpha)$, for almost all times t.
At isolated times $t_c$, the leading order coefficient is not necessarily
$-2$. It can be of the form $-g(g+1)$ (see below), but the exponent of the
leading term is always $-2$. 

Hence, an elliptic function ansatz for $u(x,t)$ can only have second order
poles and with the substitution $u(x,t)=2 \p_x^2 \ln \vartheta(x,t)$ gives
rise to a \weier expansion of the form \rf{facthetaper}, with an arbitrary
prefactor $\exp(c(x,t))$, for an arbitrary function $c(x,t)$, entire in $x$. 
Substitution of this ansatz in the KdV equation then determines that $c_{xx}$
is  doubly periodic and meromorphic in $x$. The only such $c$ is a constant.
Hence, all finite-gap elliptic solutions of the KdV equation are of the form
\rf{sol}. 

\subsection{If $L_1/L_2\gg 1$, any number of poles $N\neq 2$ in the
fundamental domain is allowed}

We have already argued that the equations (\ref{usolhyp}) and
(\ref{soldynsys}-b) are obtained from \rf{sol} and (\ref{eldynsys}-b) in the
limit $L_1\rightarrow \infty$.  On the other hand, \rf{weierpois} can be
regarded as a perturbative expansion of the soliton case, which corresponds to
its first two terms. Because \rf{weierpois} converges uniformly away from
$x=n_1L_1+i n_2 L_2$, for arbitrary integers $n_1, n_2$ we conclude that for
large (but finite) values of $L_1/L_2$ any real, nonsingular $g$-soliton
solution with rationally related wave numbers 
$(k_1:k_2:\ldots:k_g)=(p_1:p_2:\ldots:p_g)$ has an elliptic deformation with
real period $L_1$ and imaginary period $i L_2$, which is real and
nonsingular\footnote{That this deformation is nonsingular follows from the
fact that the limit of such a deformation is the original soliton solution:
the limit of a singular solution results in a singular soliton solution. This
is impossible, because only nonsingular soliton solutions are considered,
hence the elliptic deformations of nonsingular soliton solutions are
nonsingular}. Since $N=p_1+p_2+\dots+p_g\neq 2$ is arbitrary in the soliton
case, this is also true for these elliptic deformations of the solitons. Hence
the constraint \rf{elcons} is solvable for arbitrary $N$, for $L_1\gg L_2$.

At this point, it is appropriate to remark that if one is interested in
elliptic solutions of the Kadomtsev-Petviashvili (KP) equation,

\beq\la{kp}
\p_x\left(-u_t+6 u u_x+u_{xxx}\right)+3 \sigma^2  u_{yy}=0
\eeq

\no (with parameter $\sigma$), then the equations \rf{sol}, (\ref{eldynsys}-b)
are replaced by \cite{chud}

\alpheqn

\bea\la{kp1}
u(x,y,t)&=&-2 \sum_{n=1}^N\wp(x-x_n(y,t)),\\
\pp{x_n}{t}&=&3 \sigma^2 \left(\pp{x_n}{y}\right)^2+12 \sum_{j=1, j\neq
n}^N\wp(x_j-x_n),\\
\sigma^2\ppn{2}{x_n}{y}&=&- 16 \sum_{j=1, j\neq
n}^N\wp'(x_j-x_n).
\eea

\resetalpheqn

\no This clarifies the appearance of the constraint \rf{elcons} on the motion
of the poles of elliptic solutions of the KdV equation, where the poles are
independent of $y$. The search for $y$-independent solutions of the KP
equation reduces it to the KdV equation and it reduces the equations
(\ref{kp1}-c) to \rf{sol}, (\ref{eldynsys}-b), forcing the poles to remain on
the invariant manifold defined by \rf{elcons}. For the KP equation, no such
constraint exists and hence the number of poles $N$ in the fundamental domain
can be any integer, not equal to two.

\subsection{For any $N>4$, nonequivalent configurations exist, for $L_1/L_2$
sufficiently large}

Again, we only consider solutions with nonsingular soliton limits; $i.e.,$ the
elliptic deformations mentioned above. Consider the asymptotic behavior for
$t\rightarrow -\infty$ of $ \lim_{L_1 \rightarrow \infty} u(x,t)$. In this
soliton limit, as $t \rightarrow -\infty$, the poles are collected in groups
corresponding to one-soliton solutions. 

In this section, two configurations are called {\em nonequivalent} if the above
asymptotic behavior results in a different grouping of the poles. 

For $N=3$, all configurations are equivalent to one configuration. In the
limit $L_1\rightarrow \infty$, this configuration corresponds to the
two-soliton case with $k_1:k_2=2:1$.  This configuration is discussed in
Section \ref{sec:dubnov}.

For $N=4$, all configurations are again equivalent to one configuration. This
configuration corresponds to the two-soliton case with $k_1:k_2=3:1$. Recall
that $k_1$ and $k_2$ are not allowed to be equal, hence a configuration with
two poles to the left and two poles to the right does not exist. Another way
of expressing that only one configuration exists is that $N=4$ can only be
decomposed as the sum of distinct positive integers without common factor as
$N=3+1$. Again, all $N=4$ configurations are equivalent. This configuration is
discussed in Section \ref{sec:4poles}. That section also discusses another
example of an $N=4$ potential which does not have a nonsingular soliton limit.
This potential is a special case of one of the Treibich-Verdier potentials
\rf{tv}. 

Any integer $N>4$ can be written as a sum of distinct positive integers
without overall common factor in more than one way\footnote{$5=1+4=2+3$,
$N=1+(N-1)=1+2+(N-3)$, for $N>5$}. Let the number of terms in the $m$-th
decomposition of $N$ be denoted as $N_m$, then $N=\sum_{i=1}^{N_m}n_i$, with
the $n_i$ distinct and having no overall common factor. This configuration
corresponds to the $N_m$-soliton case with wavenumber ratios $(k_1:k_2:\ldots
:k_{N_m})=(n_1:n_2:\ldots :n_{N_m})$. A solution with these wave numbers has
$N_m$ phases and is an $N_m$-soliton solution. Hence for any $N>4$ there exist
at least as many different configurations as there are decompositions of $N$
into distinct positive integers, without overall common factor. These
configurations need not have the same number of phases.

Two nonequivalent configurations corresponding to $N=5$ are discussed in
Section \ref{sec:n=5}. 

\subsection{The poles only collide in triangular numbers}

A collision of poles is a local process in which only the colliding poles play
a significant role. The analysis of the collisions is identical to the
rational and the soliton cases because close to the collision point, the
\weier function reduces to $1/x^2$. Kruskal \cite{kruskalpoles} already
noticed that the poles do not collide in pairs, but triple collisions do
occur. In fact, any triangular number of poles can participate in a collision,
in which case the solution of the KdV equation at the collision time $t_c$,
nearby the collision point $x_c$ is given by $u(x,t_c)=-g(g+1)/(x-x_c)^2$
\cite{amm}. Asymtotically near the collision point $x_c$, $i.e.,$ $t<t_c$, the
poles lie on the vertices of a regular polygon with $g(g+1)/2$ vertices. For
$t>t_c$, the poles emanate from the collision points, again forming a regular
polygon with $g(g+1)/2$ vertices. If $g(g+1)/2$ is even, this polygon is
identical to the polygon before the collision. If $g(g+1)/2$ is odd, the
polygon is rotated around the collision point by $2 \pi/g(g+1)$ radians. 

Of all these collision types, the one where three poles collide (corresponding
to $g=1$) is generic. It is the one observed in the examples illustrated in
Section \ref{sec:ex}. 

Since the poles only collide in triangular numbers, it is possible that at any
given time $t_c$ the solution of the KdV equation has the form \rf{gw}, with
not all $g_i=1$. At almost every other time $t$, such a solution has
$N=\sum_{i=1}^M g_i(g_i+1)/2$ distinct poles. 

\subsection{The solutions \rf{sol} are finite-gap potentials of the stationary
Schr\"{o}dinger equation \rf{schrodinger}}

By construction the solutions \rf{sol} are periodic in $x$ because they are
obtained as a \weier factorization of the theta function appearing in
\rf{thetasol}, upon which we have imposed the double periodicity. Hence the
solutions \rf{sol} are finite-gap potentials of the Schr\"{o}dinger
operator. In \cite{gw2}, another proof of this can be found. 

For solutions that are elliptic deformations of the nonsingular solitons of
Section \ref{sec:thick}, more can be said: an elliptic deformation of a
$g$-soliton solution is a $g$-gap potential of the Schr\"{o}dinger equation.
The reasoning is as follows: we already know that any elliptic deformation
results in a finite-gap potential of the Schr\"{o}dinger equation. On the
other hand, any finite gap potential of the Schr\"{o}dinger equation is of the
form \rf{thetasol}. The soliton limit of such a finite-gap solution with
$g$-phases is a $g$-soliton solution \cite{belokolos1}. Hence the number of
phases of an elliptic deformation of a $g$-soliton solution is equal to $g$. 

This limit is the soliton limit of the periodic solutions, in which the
fundamental domain reduces to the fundamental strip. In order to have a
$g$-soliton solution of the KdV equation, we remarked in Section
\ref{sec:thick} that at least $N=g(g+1)/2=1+2+\ldots+g$ poles are required in
the fundamental strip. Hence, this many poles are required in the fundamental
domain to obtain a $g$-gap potential of the Schr\"{o}dinger equation that is
an elliptic deformation of a nonsingular soliton solution. 

\section{Examples}\la{sec:ex}

In this section, some explicit examples of elliptic solutions of the KdV
equation are discussed. These are illustrated with figures displaying the
motion of the poles in the fundamental domain. Other figures display the
solution of the KdV equation $u(x,t)$ as a function of $x$ and $t$. All these
figures were obtained from numerical solutions of the corresponding
constrained dynamical system. In all cases, the constrained dynamical system
was solved using a projection method: the dynamical system \rf{eldynsys} is
used to evolve the system for some time. Subsequently, the new solution is
projected onto the constraints \rf{elcons} to correct numerical errors, after
which the process repeats. 

In all examples given, $L_1=4$ and $L_2=\pi$. This seems to indicate that one
can wander far away from $L_1/L_2\gg 1$ and still obtain soliton-like elliptic
solutions of the KdV equation. This is not surprising as
\rf{weierpois} indicates that as perturbation parameter on the soliton case
one should use $\epsilon=\exp(-2 \pi L_1/L_2)$. For the values given above,
this gives $\epsilon=0.00034$. 

\subsection{The solution of Dubrovin and Novikov \cite{dubnov}:
$\mbf{N=3}$}\la{sec:dubnov}

\begin{figure} 
\begin{tabular}{cc} 
\psfig{file=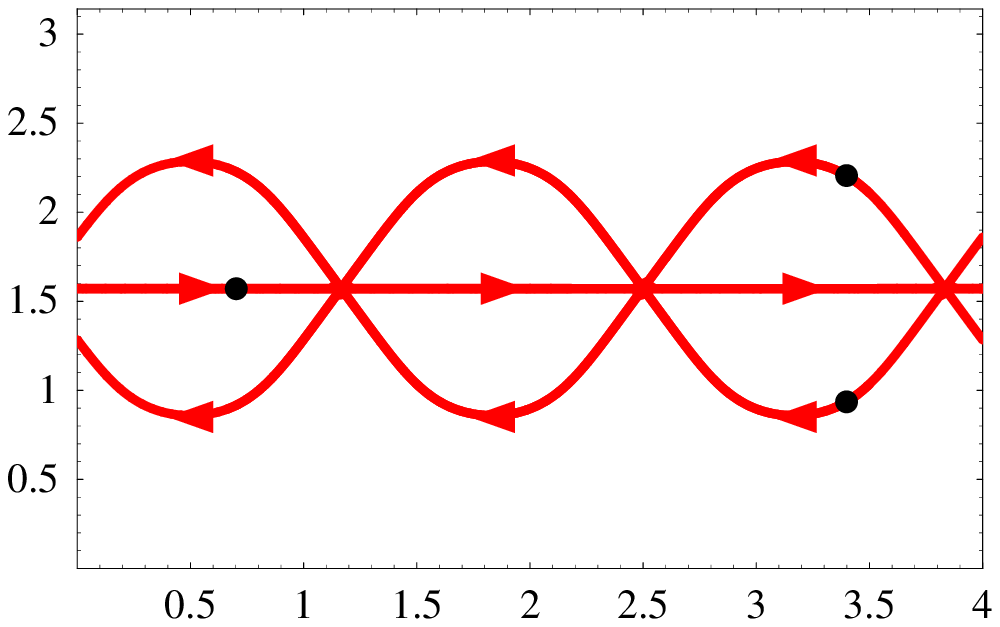,width=3in}&
\psfig{file=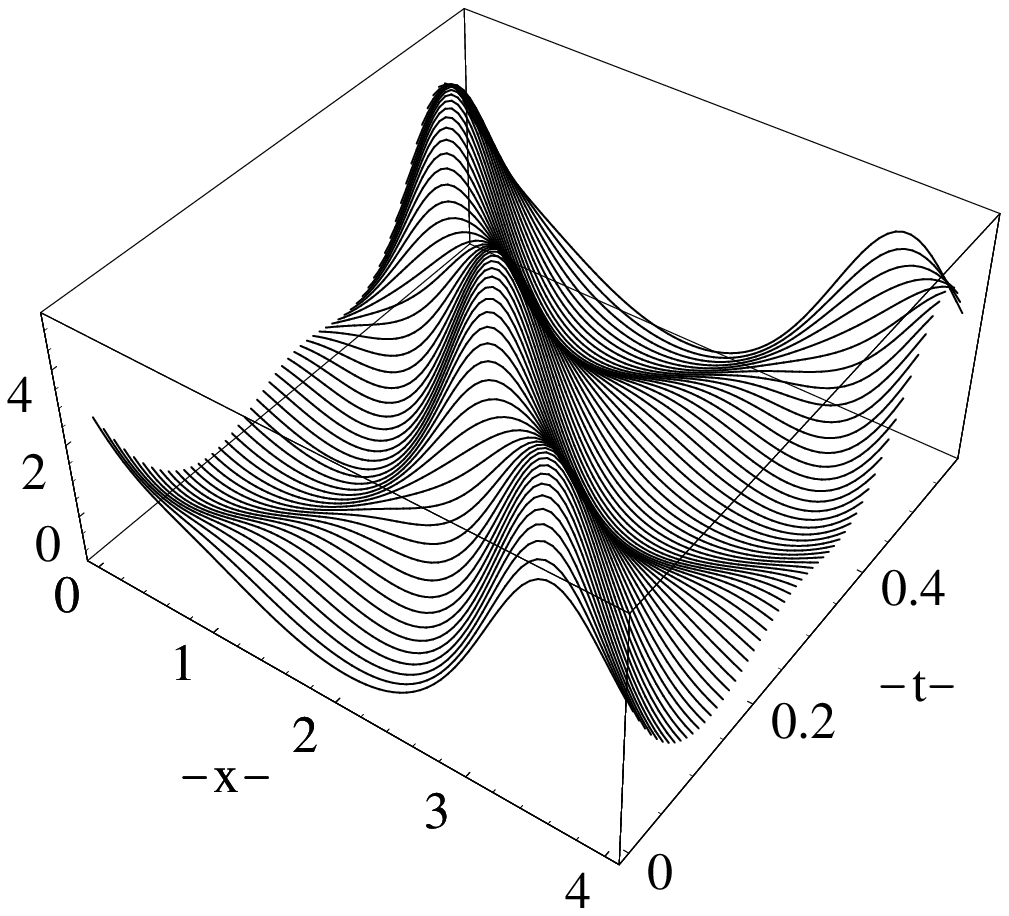,width=3in}\\ 
(a)&(b) 
\end{tabular}
\caption{\label{fig3} {\bf The solution of Dubrovin and Novikov, with $L_1=4$
and $L_2=\pi$. (a) The motion of the poles in the fundamental domain. The
initial position of the poles is indicated by the black dots. The arrows
denote the motion of the poles. (b) The KdV solution $u(x,t)$}} 
\end{figure}

Dubrovin and Novikov \cite{dubnov} integrated the KdV equation with  the
Lam\'{e}-Ince potential $u(x,0)=-6\wp(x-x_c)$ as initial condition. They found
the solution to be elliptic for all time, with $N=3$. They gave explicit
formulae for the solution, which they remarked was probably the simplest
two-gap solution of the KdV equation. The dynamics of the poles in the
fundamental domain is displayed in Fig. \ref{fig3}a. Fig. \ref{fig3}b displays
the corresponding two-phase solution of the KdV equation. Animations of the
behavior of the poles and of $u(x,t)$ as $t$ changes are also available at
{\tt http://amath-www.colorado.edu/appm/other/kp/papers}. Notice the
soliton-like interactions of the two phases in the solution. In terms of the
classification of Lax \cite{lax2}, these are interactions of type (c) ($i.e.,$
$u(x,t)$ has only one maximum while the larger wave overtakes the smaller
wave). 

From Fig. \ref{fig3}a, it appears that the Dubrovin-Novikov solution is
periodic in time. This was indeed proven by \`{E}nol'skii \cite{enolskii}. 

For this specific solution only one of the three constraint equations is
independent: since the derivative of the \weier function is odd, the sum of the
constraints is zero. Furthermore, labelling the three poles by $x_1, x_2$ and
$x_3$, for reality $x_2=x_1^*+i L_2$ and $x_3$ is on the centerline. Hence the
second constraint is the complex conjugate of the first constraint. The
constraints \rf{elcons} reduce to the single equation

\beq\la{dubnovcons}
\wp'(x_1-x_1^*)+\wp'(x_1-x_3)=0.
\eeq

\no This equation was solved numerically to provide the initial condition
shown in Fig. \ref{fig3}a. The initial guess required for the application of
Newton's method is based on the knowledge of the soliton limit. In that case
two poles on the right represent a faster soliton, one pole on the left
represents the slower soliton. The periodic case is not that different: the
vertical line of poles with the smallest vertical distance between poles has
poles closer to the real $x$-axis than the others and correspond to the wave
crest with the highest amplitude, as seen in Fig. \ref{fig3}b. We refer to the
Dubrovin-Novikov solution as a $(2:1)$-solution because of the natural
separation of the poles in a group of 2 poles $(x_1,x_2)$ and a single pole
$(x_3)$. 

Equating $x_1=x_3+\epsilon$ and only condering the singular terms of
\rf{dubnovcons}, it is possible to examine the location of the poles close to
a collision points $x_c$. With $\wp'(x)=-2/x^3$ in this limit and
$\epsilon=\epsilon_r+i \epsilon_i$, one finds

\beq\la{closetocol}
\epsilon_r^3-3 \epsilon_i^2\epsilon_r=0,~~~~ 9 \epsilon_i^3-3 \epsilon_r^2
\epsilon_i=0.
\eeq

\no This set of equations has three solutions, corresponding to the three
distances between the poles: $\epsilon_r/\epsilon_i\in\left\{0, \sqrt{3},
-\sqrt{3}\right\}$. This allows for two triangular configurations of the
poles: an equilateral triangle pointing left of the collision point and one
pointing right. 

Using the dynamical system \rf{eldynsys} in the same way and only retaining
singular terms results in 

\beq\la{dubnovdynsys} \dot{\epsilon}_r=-\frac{3}{4 \epsilon_i^2}+3
\frac{\epsilon_r^2-\epsilon_i^2}{\left(\epsilon_r^2+\epsilon_i^2\right)^2},~~~~
\dot{\epsilon}_i=\frac{6 \epsilon_r
\epsilon_i}{\left(\epsilon_r^2+\epsilon_i^2\right)^2}. \eeq

\no Since the constraints \rf{elcons} are invariant under the flow
\rf{eldynsys}, the solutions to \rf{closetocol} give invariant directions of
the system \rf{dubnovdynsys}. Along these invariant directions, one obtains
ordinary differential equations for the motion of the poles as they approach
the collision point. It follows from these equations that the poles approach
the collision point $x_c$ with infinite velocity. Integrating the equations
with initial condition $\epsilon(t_c)=0$ gives 

\beq\la{scalinglaw}
\epsilon=\frac{3^{5/6}}{2}(\sqrt{3}+i)(t_c-t)^{1/3}.
\eeq

\no Using the three branches of $(t_c-t)^{1/3}$ results in the dynamics of
each edge of the equilateral triangle. If $t<t_c$ this triangle is pointing
left, for $t>t_c$ it is pointing right. 

\subsection{$\mbf{N=4}$: an elliptic deformation and a Treibich-Verdier
solution}\la{sec:4poles}

The next solution we discuss has 4 poles in the fundamental domain and is an
elliptic deformation of a soliton solution. In the limit $L_1\rightarrow
\infty$, this solution corresponds to a two-soliton solution with wavenumber
ratio $k_1/k_2=3/1$, so this solution is refered to as a $(3:1)$- solution.

\begin{figure}[ht] \centerline{\psfig{file=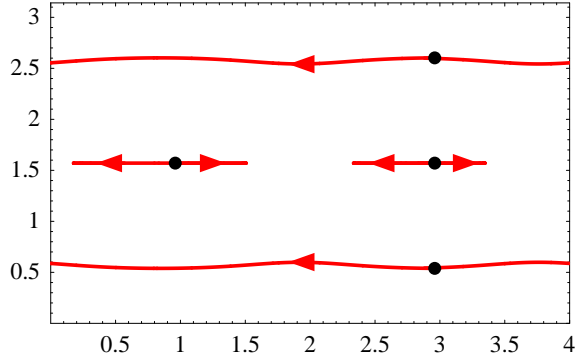,width=3in}} 
\vspace*{-0.5in} \caption{\label{fig5} {\bf $N=4$: the pole dynamics of a
$(3:1)$-solution. The initial positions of the 4 poles are indicated. The
arrows on the centerline indicate that the poles there move in both
directions.}} \end{figure}

The motion of the poles in the fundamental domain is displayed in Fig.
\ref{fig5}. Corresponding to the given wavenumber ratio, the amplitute ratio
of the two phases present in the solution is roughly $k_1^2/k_2^2=9/1$. As a
consequence, the form of $u(x,t)$ is not very illuminating and it has been
omitted. Animations with the time dependence of both the positions of the
poles and of $u(x,t)$ are again available at {\tt
http://amath-www.colorado.edu/appm/other/kp/papers}. 

Note that the poles of the $(3:1)$-solution do not collide. This is in
agreement with the results of Thickstun \cite{thickstun} who outlined which
configurations lead to collisions and which do not, in the hyperbolic case. As
mentioned before, the examination of collision behavior is essentially local
and no differences appear between the rational, hyperbolic and elliptic cases. 

Another configuration with $N=4$ exists. Consider the potential

\beq\la{tv4}
u(x,t=t_c)=-2 \wp(x-x_0)-6 \wp(x-x_0-\omega_1),
\eeq

\no with $x_0$ on the centerline. This is a Treibich-Verdier potential,
obtained from \rf{gw} with $M=2$, $g_1=1$, $g_2=2$, $\alpha_1=x_0$ and
$\alpha_2=x_0+\omega_1$. It is referred to as a Treibich-Verdier potential
because the position of the poles is given in terms of the periods of the
\weier function, as in \rf{tv}. Also, it can be obtained from \rf{tv} as a
degenerate case. As before $N=\sum_{i=1}^{2}g_i(g_i+1)/2=4$, hence for all
times that are not collision times, this solution has 4 distinct poles in the
fundamental domain. The time $t=t_c$ is a collision time. Immediately after
the collision time $t=t_c$, the 3 poles located at $x_0+\omega_1$ separate, as
in the Dubrovin-Novikov solution, along an equilateral triangle. The result
appears to be a three-phase solution. However, it is known that the potential
\rf{tv4} is a two-gap potential of the Schr\"{o}dinger equation and its
hyperelliptic Riemann surface is given explicitly in \cite{belokolos1}. This
solution is not an elliptic deformation of a nonsingular soliton solution and
the separation into different phases does not make sense. This is also seen
from the following argument: if, for a fixed time which is not a collision
time, we attempt to take the limit as $L_1=2 \omega_1\rightarrow \infty$,  the
poles seem to separate in three distinct solitons with respective wave numbers
$(k_1:k_2:k_3)=(2:1:1)$. Such a nonsingular soliton solution does not exist
for the KdV equation and the separation into different phases does not make
sense. 

\begin{figure}  \begin{tabular}{cc} 
\psfig{file=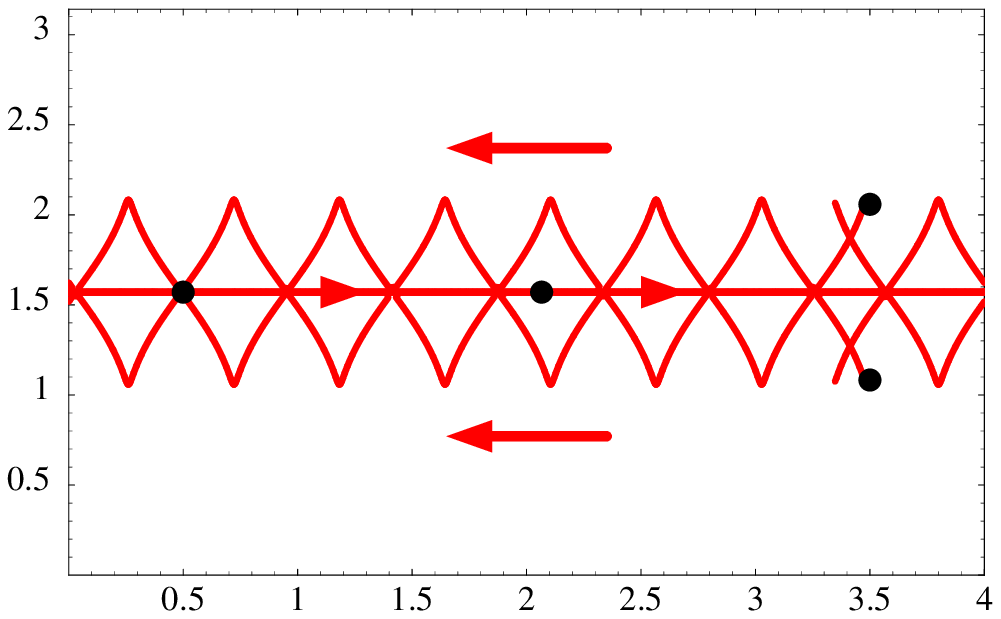,width=3in}&
\psfig{file=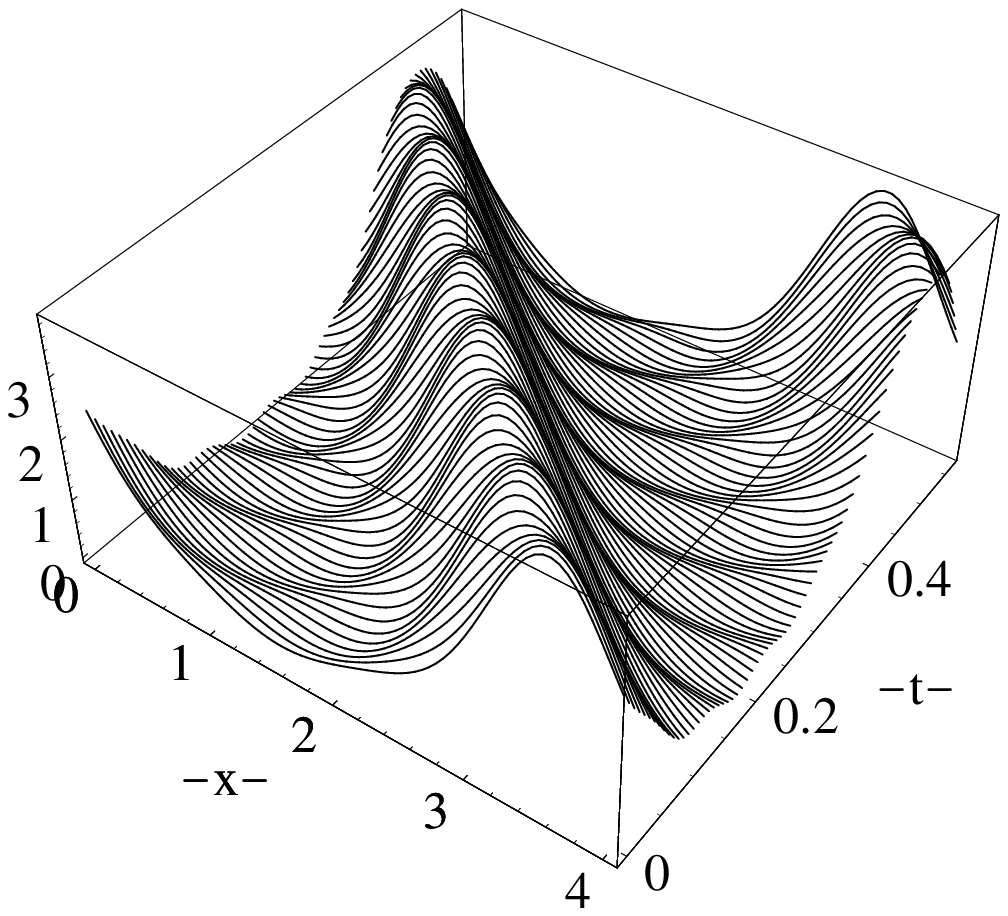,width=3in}\\  (a)&(b)  \end{tabular}
\caption{\label{figtv4} {\bf An $N=4$, $M=2$ Treibich-Verdier solution,  with
$L_1=4$ and $L_2=\pi$.  (a) The motion of the poles in the fundamental domain.
The initial position of the poles is indicated by the black dots. The initial
time $t=0$ was chosen different from the collision times $t_c$. The arrows
denote the motion of the poles. (b) The KdV solution $u(x,t)$}}  \end{figure}

The dynamics of the poles is illustrated in Fig. \ref{figtv4}a. The
corresponding KdV solution is shown in Fig. \ref{figtv4}b. 

The dynamics of the poles illustrated in Fig. \ref{figtv4}a exhibits behavior
that appears qualitatively different from any other solution discussed here.
The trajectories traced out by the motion of the poles in the fundamental
domain appear to have singular points (cusps), away from the collision points.
Upon closer investigation, these ``cusps'' are only a figment of the
resolution of the plot and the poles trace out a regular curve as a function
of time, away from the collision times. Exactly why the global pole dynamics
of the Treibich-Verdier potential \rf{tv4} under the KdV flow appears so
different from the pole dynamics of elliptic deformations of soliton solutions
of the KdV equation is an open problem. Another question one may ask is
whether similar behavior is observed for other solutions originating from
Treibich-Verdier potentials. 

\subsection{$\mbf{N=5}$: two different possibilities}\la{sec:n=5}

For $N=5$, two soliton configurations are possible, and corresponding to each
of these is an elliptic solution. The first solution is a $(4:1)$-solution.
The second solution is a $(3:2)$-solution. 

\begin{figure}[p] 
\begin{tabular}{cc}
\multicolumn{2}{c}{\psfig{file=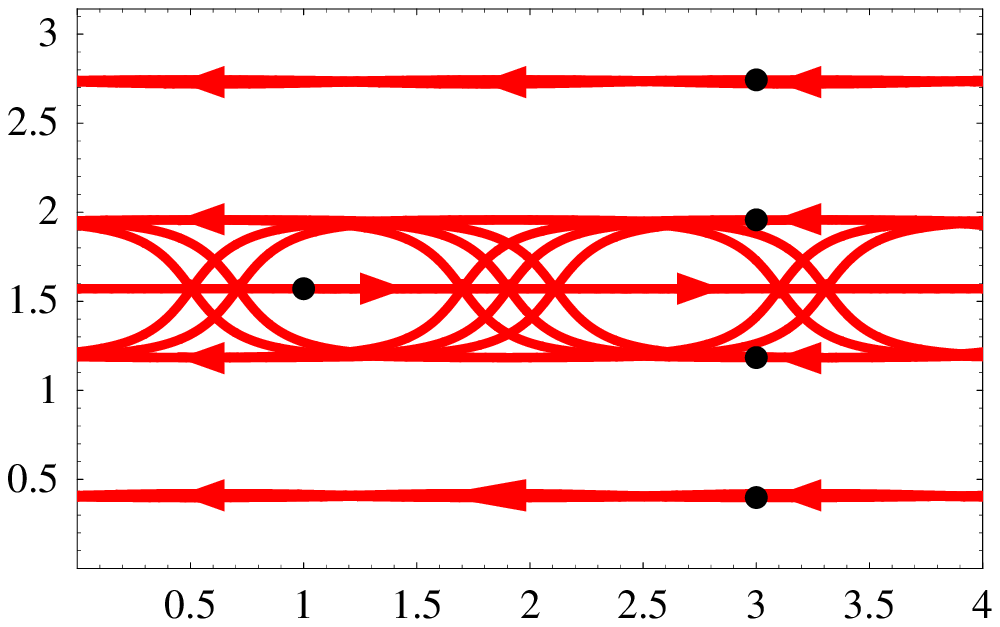,width=3in}}\\
\multicolumn{2}{c}{(a)}\\ 
\psfig{file=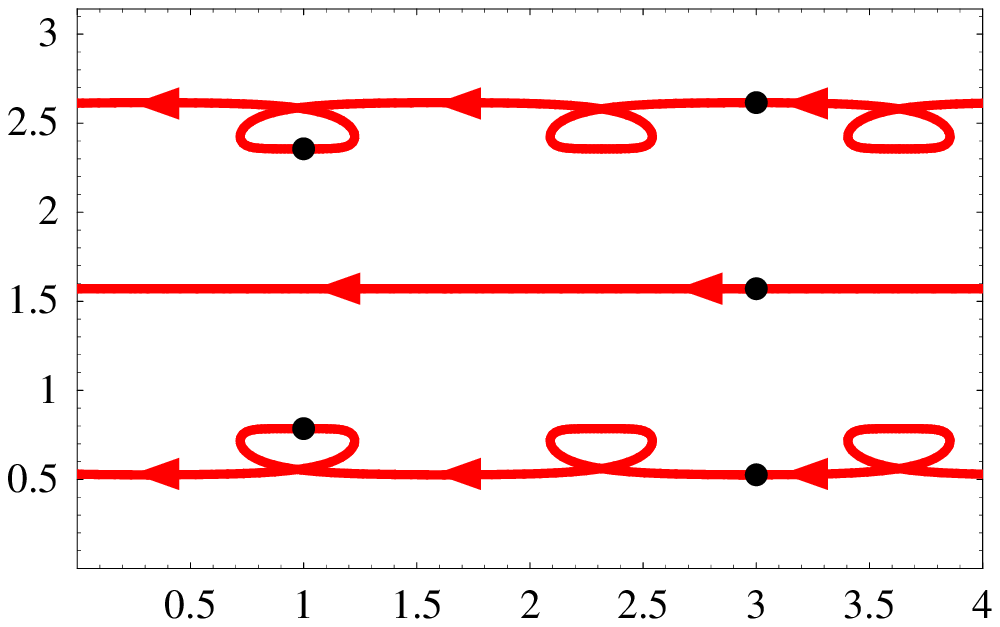,width=3in}&
\psfig{file=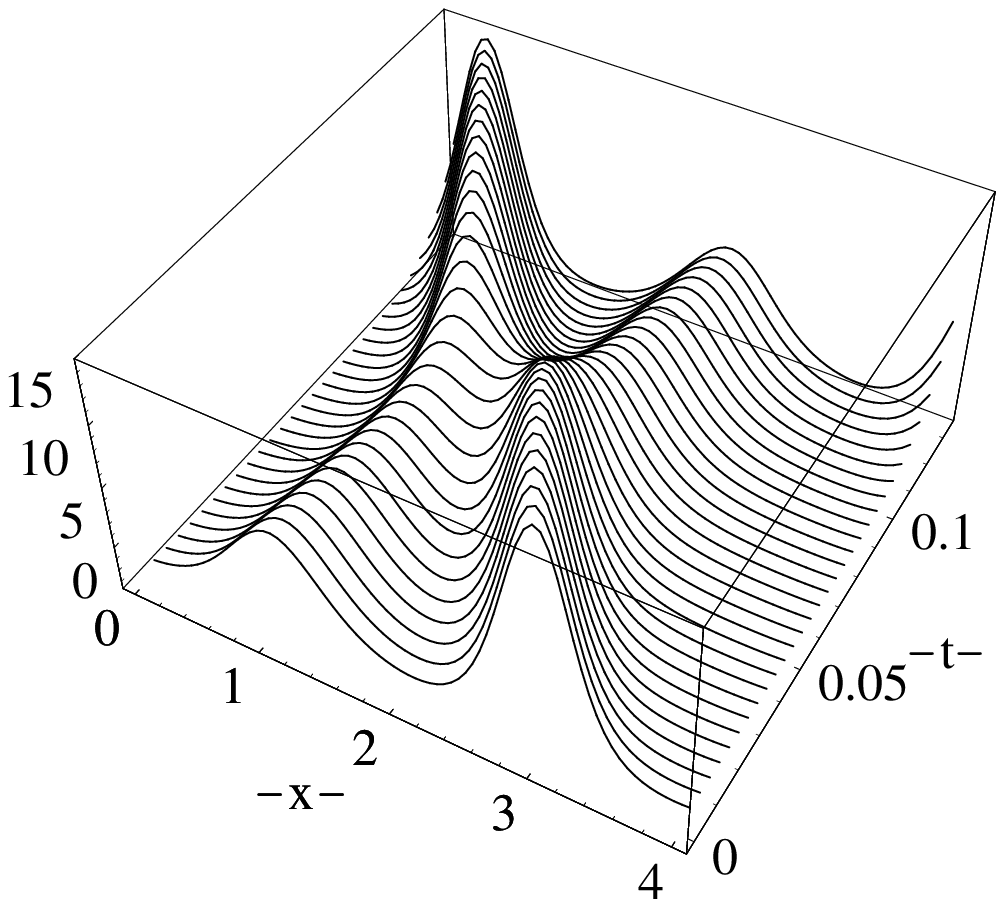,width=3in}\\ (b)&(c) \end{tabular}
\caption{\label{fig6} {\bf $N=5$: (a) The pole dynamics of a $(4:1)$-solution in
the fundamental domain. 
(b) The pole dynamics of a $(3:2)$-solution in the fundamental domain.
(c) The KdV solution $u(x,t)$ corresponding to the pole dynamics in (b). In (a)
and (b),
the initial positions of the poles are indicated. Both solutions are
quasiperiodic in time. 
$u(x,t)$}} \end{figure}

The $(4:1)$-solution offers no new pole-dynamics: initially 1 pole is located
on the centerline, at the left in the fundamental domain. The other poles are
located at the right of the fundamental domain, symmetric with respect to the
centerline. The three poles closest to the centerline interact as the
$(2:1)$-solution. The two outer poles behave as the two outer poles of the
$(3:1)$-solution. The pole dynamics of the $(4:1)$-solution is displayed in Fig.
\ref{fig6}a.

The $(3:2)$-solution is more interesting. It is displayed in Fig. \ref{fig6}c,
together with the motion of the poles in the fundamental domain \ref{fig6}b.
Again, the two crests of $u(x,t)$ interact in a soliton-like manner. In Lax's
classification \cite{lax2}, this is an interaction of type (a), where at every
time two maxima are observed. Fig. \ref{fig6}b only displays the motion of the
poles for a short time, in order not to clutter the picture. The motion of the
poles is presumably quasiperiodic in time, as is the case for the
$(4:1)$-solution. It appears that the two poles above (or below) the middle
line of the fundamental domain share a common trajectory. It is an open
problem to establish whether or not this is the case. 

\subsection{$\mbf{N=6}$: two different possibilities. A three-phase solution}

For $N=6$, two distinct pole configurations are possible. The first one
corresponds to a $(5:1)$-solution and results in a two-gap potential of the
Schr\"{o}dinger equation. It essentially behaves as the $(3:1)$-solution with
two more poles added, which also behave as the outer poles of the
$(3:1)$-solution. 

The second configuration is a $(3:2:1)$-solution, which limits to a
three-soliton solution with wavenumber ratio $(k_1:k_2:k_3)=(3:2:1)$. This
elliptic solution is a three-phase solution of the KdV equation. 

\begin{figure}[p]
\centerline{\psfig{file=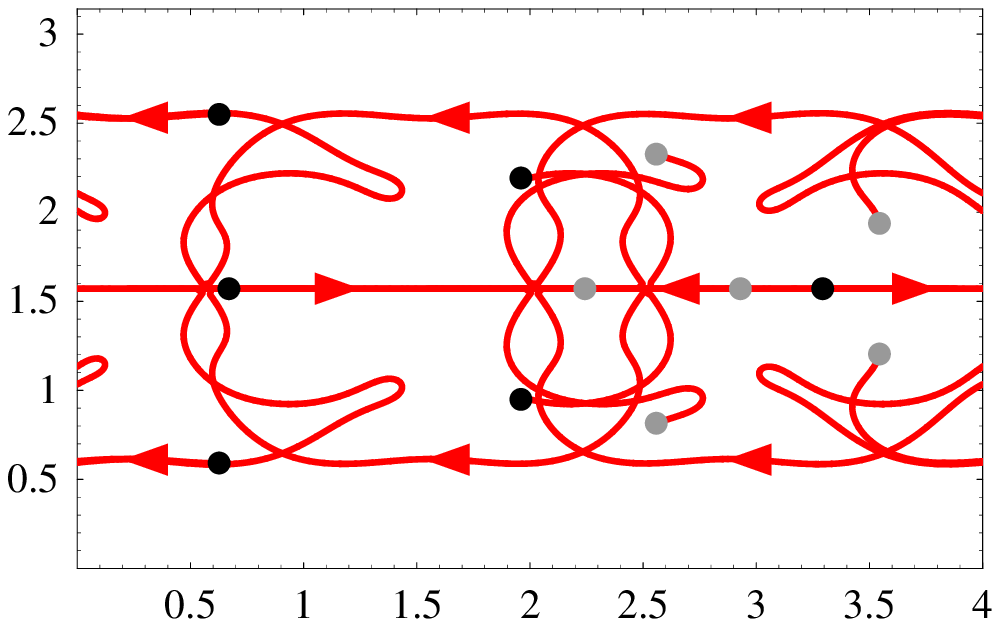,width=4in}} 
\vspace*{-0.7in}
\caption{\label{fig8} {\bf The pole dynamics of a $(3:2:1)$-solution. The black
dots mark the initial position of the poles; the grey dots mark the position of
the poles at $t=0.4$.}}
\centerline{\psfig{file=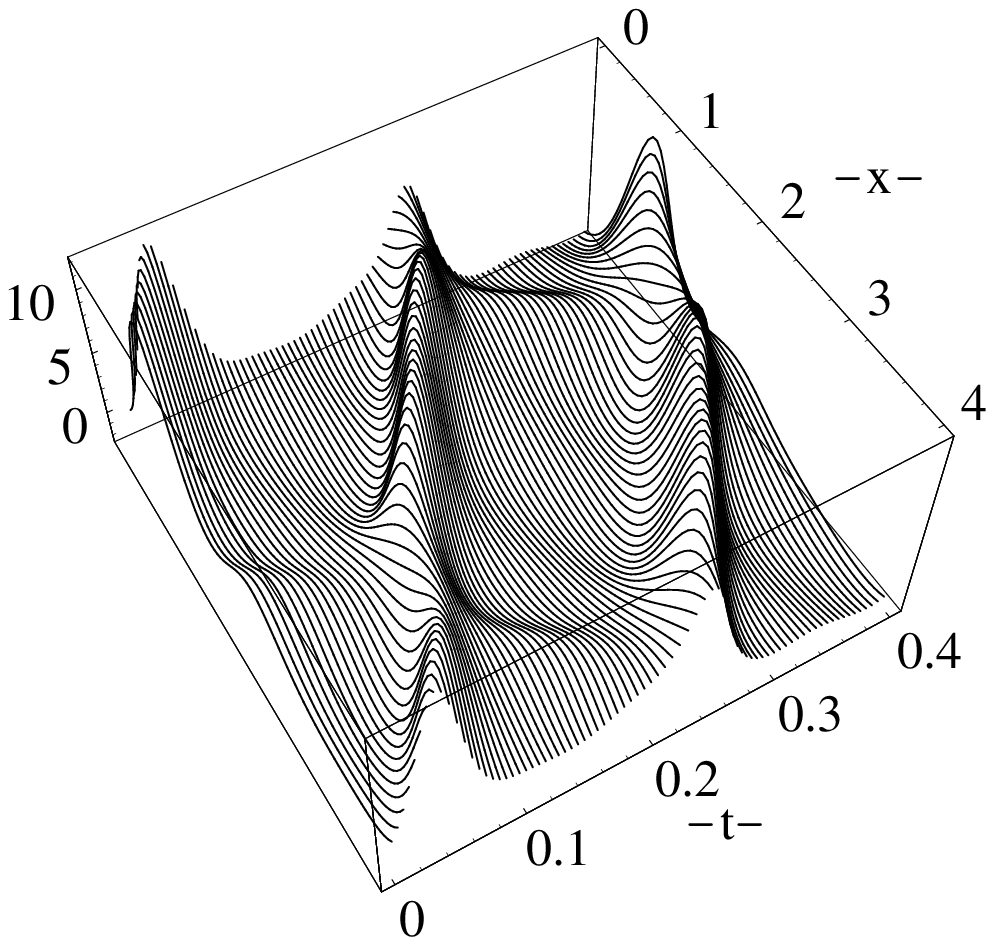,width=4in}} 
\caption{\label{fig9} {\bf A $(3:2:1)$-solution of the KdV equation,
corresponding to the pole dynamics in Fig. \ref{fig8}}}
\end{figure}

The amplitude ratio of the $(3:2:1)$-solution is $(9:4:1)$, which explains why
the third phase is hard to notice in Fig. \ref{fig9}. Animations of the pole
dynamics and of the time dependence of the $(3:2:1)$-solution are available at
{\tt http://amath-www.colorado.edu/appm/other/kp/papers}.

\section*{Acknowledgements} The authors acknowledge useful discussions with  
B. A. Dubrovin, S. P. Novikov, C. Schober and A. P. Veselov. This work was
carried out at the University of Colorado and the Mathematical Sciences
Research Institute. It was supported in part by NSF grants DMS 9731097
and DMS-9701755.

\bibliographystyle{plain}

\bibliography{}

\end{document}

%% file: macrosarticle.tex
\newcommand{\beq}{\begin{equation}}
\newcommand{\eeq}{\end{equation}}
\newcommand{\ba}{\begin{array}}
\newcommand{\ea}{\end{array}}
\newcommand{\bea}{\begin{eqnarray}}
\newcommand{\eea}{\end{eqnarray}}
\newcommand{\bc}{\begin{center}}
\newcommand{\ec}{\end{center}}
\newcommand{\bt}{\begin{table}}
\newcommand{\et}{\end{table}}

\newcommand{\la}[1]{\label{#1}}
\newcommand{\p}{\partial}

\newcommand{\no}{\noindent}
\newcommand{\pp}[2]{{\partial #1 \over \partial #2}}
\newcommand{\ppn}[3]{{\partial^{#1} #2 \over \partial #3^{#1}}}

\newcommand{\mbf}[1]{\mbox{\boldmath {$#1$}}}

\newcommand{\rf}[1]{(\ref{#1})}
\newcommand{\beqno}{\begin{displaymath}}
\newcommand{\eeqno}{\end{displaymath}}

\newcommand{\been}{\begin{enumerate}}
\newcommand{\een}{\end{enumerate}}

\renewcommand{\iff}{\Leftrightarrow}
\newcommand{\since}[1]{\stackrel{\mbox {using \rf{#1}}}{=}}

\newcommand{\weier}{Weierstra\ss~}

\newlength{\myheight}
\newlength{\mylength}

\makeatletter
\let\c@table=\c@mytable
\makeatother

\makeatletter
\let\c@figure=\c@myfigure
\makeatother

\makeatletter
\let\c@equation=\c@myequation
\makeatother

\renewcommand{\theequation}{\arabic{section}.\arabic{equation}}

\newcounter{saveeqn}

\newcommand{\alpheqn}{\setcounter{saveeqn}{\value{equation}}
\stepcounter{saveeqn}\setcounter{equation}{0}
\renewcommand{\theequation}{\mbox{\arabic{section}.\arabic{saveeqn}
\alph{equation}}}}

\newcommand{\resetalpheqn}{\setcounter{equation}{\value{saveeqn}}
\renewcommand{\theequation}{\arabic{section}.\arabic{equation}}}

\newtheorem{example}{Example}[section]